%% file: RJwrapper.tex
\documentclass[a4paper]{report}
\usepackage[utf8]{inputenc}
\usepackage[T1]{fontenc}
\usepackage{RJournal}
\usepackage{amsmath,amssymb,array}
\usepackage{booktabs}

\usepackage{longtable}
\usepackage{algorithm}
\usepackage{algpseudocode}
\usepackage{subfigure}

\graphicspath{{./figures/}}

\begin{document}

\sectionhead{Preprint}
\volume{}
\volnumber{}
\year{}
\month{}

\begin{article}
  \input{tzeng-hsu}

\end{article}

\end{document}

%% file: tzeng-hsu.tex
\title{The R Package \pkg{HCV} for  Hierarchical Clustering from Vertex-links}
\author{by ShengLi Tzeng and Hao-Yun Hsu}

\maketitle

\abstract{
The \pkg{HCV} package implements the hierarchical clustering for spatial data. It requires clustering results not only homogeneous in non-geographical features among samples but also geographically close to each other within a cluster. We modified typically used hierarchical agglomerative clustering
 algorithms to introduce the spatial homogeneity, by considering geographical locations as vertices and  converting spatial adjacency into whether a shared edge exists between a pair of vertices.  The main function \texttt{HCV} obeying constraints of the vertex links automatically enforces the spatial contiguity property at each step of iterations.
In addition, two methods to find an appropriate number of clusters and to report cluster members are also provided.}

\section{Introduction}

Clustering analysis is a practical means to explore heterogeneity among subsets of the data.
Clustering algorithms aim to partition data into groups by maximizing the within-group homogeneity and the between-group heterogeneity at the same time. When it comes to spatial data, however, we need to carefully define what to cluster. This is because spatial data have two types of attributes: geometrical attributes such as geographical coordinates, and non-geometrical features such as values of temperature and rainfall. We shall refer to attributes on the former as the geometry domain and the latter as the feature domain. 
In many applications, clustering results are required to 
take care of information from both domains, and keeping the spatial contiguity within clusters is of great interest.

What makes things more complicated is that the analytical methods for spatial data are usually taxonomized into three categories, namely, point-level or geostatistical data, areal or lattice data, and point patterns (see \citealp{banerjee2003hierarchical} or \citealp{cressie1993statistics} ). We collect areal or lattice data when a geometry domain is divided into several subregions at which non-geometrical features are aggregated.
Point-level or geostatistical data are collected at finite number of sites over the geometry domain for certain spatial random variables.  A point pattern consists of the characteristics of individual locations, their variations and interactions in the geometry domain.
Clustering of point patterns has been well developed; see \citet{kriegel2011density} for a review, and we will not consider this kind of problem in what follows.  This article proposes a method that can be applied to the other two situations in the same manner. We shall start from a brief review of existing works. 

For clustering areal data, where each spatial unit is typically represented by a polygon, the R package  \CRANpkg{ClustGeo} \citep{chavent2018clustgeo} balances the proximities in geometry domain and feature domain by a weighted sum of both. However, \pkg{ClustGeo} may find clusters not necessarily having spatial contiguity. 
To bypass this issue, 
we have to rely on the graph adjacency between vertices by treating each spatial unit as a vertex. 
For example, \citet{assunccao2006efficient} used minimum spanning trees to identify  strictly  contiguous  geographical  regions  with homogeneity on the feature domain. \citet{carvalho2009spatial} modified usual agglomerative hierarchical clustering methods to achieve a similar goal.

For clustering point-level spatial data, an often seen name  is \textit{dual clustering}, e.g., \citet{zhou2007dcad} and \citet{lin2005dual}. It uses k-means or k-medoids clustering accounting for spatial proximities, but the results sometimes  lead to spatially fragmented clusters. \citet{liao2012clustering} handled the issue with local search, and \citet{xie2019defining} introduced constraints via Delaunay triangulations \citep{chew1989constrained}. The R package \CRANpkg{SPODT} \citep{gaudart2015spodt} utilized  oblique regression trees to partition the geometry domain.

Although clustering of spatial data has many methodological and practical applications, 
methods for different categories of data types seem to be developed independently. Very few  point-level clustering methods can be directly applied to areal data, while algorithms for areal data rarely consider point-level cases.
In this article, we  present a package \pkg{HCV} in R that implements an algorithm to deal with data in both types in a unified way. Methods for finding an appropriate number of clusters are also provided.

The rest of the paper is organized as follows. In the next section, we explain the required input formats and key subroutines in \pkg{HCV}. Next, we demonstrate the implemented functions in this package using simulated data and a real social-economic dataset. Finally, Section Summary concludes the paper.


\section{An overview of \pkg{HCV} functionality}

We implemented a function called \texttt{HCV} as the fundamental
agglomerative clustering for areal data, which was also discussed in \citet{carvalho2009spatial}, taking into account the constraints of vertex links on the geometry domain. We will demonstrate that such agglomerative clustering can be also applied to point-level data, through a intermediate tessellation. Therefore, we can analyze data in both types with an identical framework. The resulting clusters are guaranteed to be spatially contiguous.  

An important subject for practices of clustering is to determine the appropriate number of clusters. The package also provides a function called \texttt{getCluster} with two methods to deal with the problem.

\subsection{Data input structures}

As mentioned above, spatial data clustering involves two domains. We describe the required input formats for areal and point-level data in the following.

\begin{figure}
\centering
\includegraphics[width=1\textwidth]{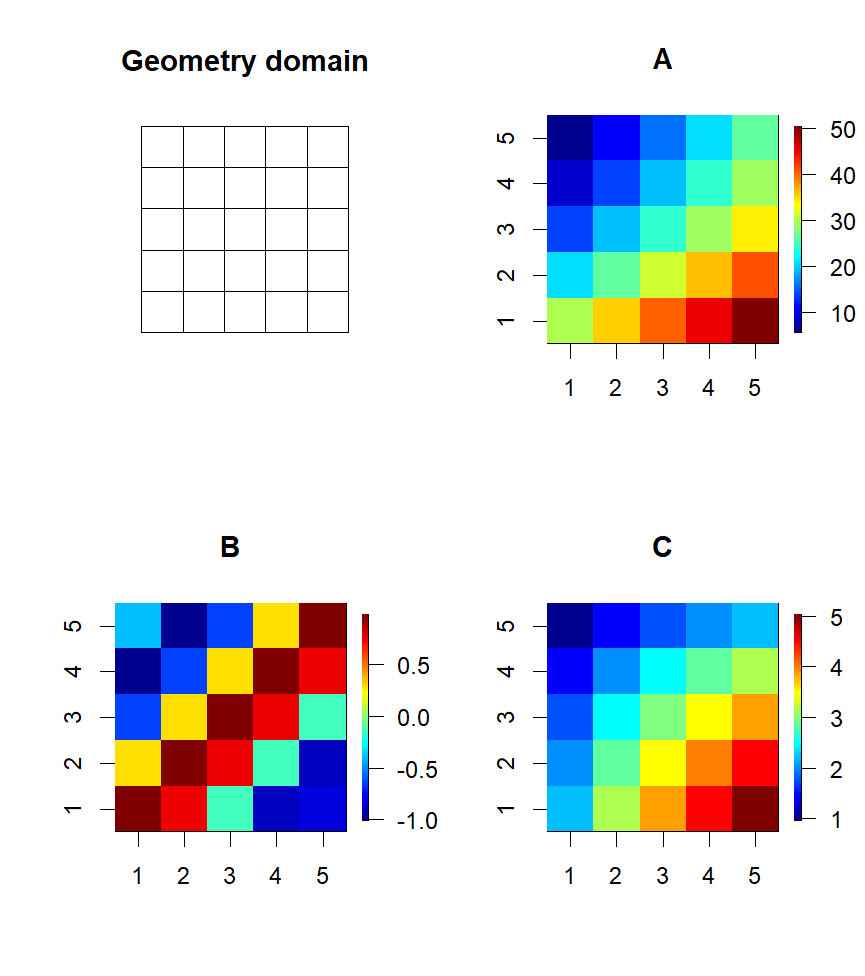}
\caption{A demonstration of inputs for areal data with  a geometry domain (top-left) and three attributes on the feature domain.}
\label{a-input}
\end{figure}

For areal data, we need to prepare a \texttt{data.frame} object whose columns stand for attributes on the feature domain, and a  \texttt{SpatialPolygonsDataFrame} object for defining polygons on the geometry domain. Here are simple codes to generate the two required objects, and the results are shown in Figure \ref{a-input}. Note that the names of polygons should match the row names of the  feature domain.
\begin{example}
library(sp)
grd <- GridTopology(c(1,1), c(1,1), c(5,5))
polys <- as(grd, "SpatialPolygons")
centroids <- coordinates(polys)
gdomain <- SpatialPolygonsDataFrame(polys,
           data=data.frame(x=centroids[,1], y=centroids[,2],
           row.names=row.names(polys)))    
fdomain <- data.frame(A=gdomain$x*5+gdomain$y^2,
                      B=cos(gdomain$x+gdomain$y),
                      C=sqrt(gdomain$x*gdomain$y))
\end{example}

\noindent The input format of geometry domain can be a adjacency matrix instead, which can be easily converted from a  \texttt{SpatialPolygonsDataFrame} object with \texttt{gTouches} in the package \pkg{rgeos} like this:
\begin{example}
adj_mat <- rgeos::gTouches(gdomain,byid=TRUE)
\end{example}

\begin{figure}
\centering
\includegraphics[width=1\textwidth]{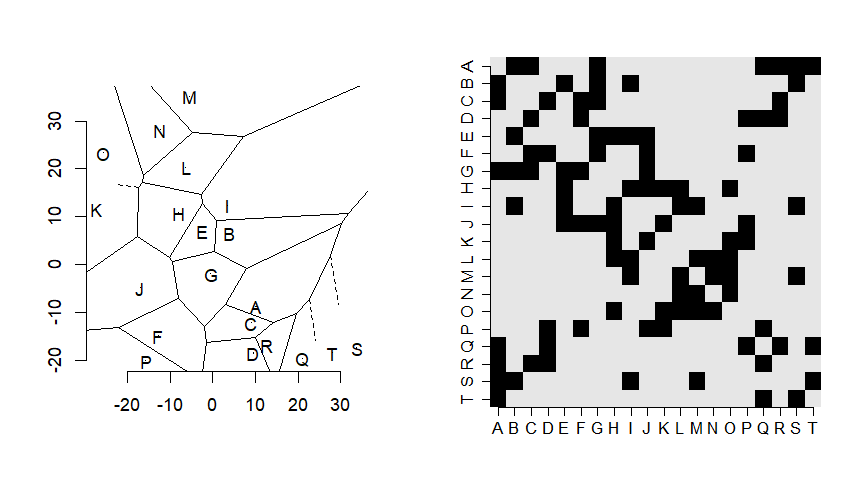}
\caption{A demonstration of a tessellation for point-level data (left) and a heatmap of the resulting adjacent matrix, where the value of black squares is 1 (right).}
\label{p-input}
\end{figure}

For point-level data, two \texttt{data.frame} object are required for the feature domain and the geometry domain. Each row of the geometry domain object is the geographical coordinate for a certain sample.  We can convert these coordinates into a adjacency matrix based on a function \texttt{tessellation\_adjacency\_matrix}. The adjacency matrix relies on Voronoi tessellation of the geometry domain, such that two samples are adjacent if their Voronoi cells share a common edge.  Below is an example of the conversion, and  Figure \ref{p-input} visualizes the result. For locations on a three-dimensional geometry domain, a Delaunay tessellation can be applied similarly. The function \texttt{tessellation\_adjacency\_matrix} will give the adjacency matrix by automatically using a suitable tessellation.
\begin{example}
library(fields)
library(alphahull)
pts <- ChicagoO3$x
rownames(pts) <- LETTERS[1:20]
Vcells <- delvor(pts)
plot(Vcells,wlines='vor',pch='.')
text(pts,rownames(pts))
Amat <- tessellation_adjacency_matrix(pts)
\end{example}

\subsection{Key subroutine: HCV algorithm}
Before introducing the fundamental  \texttt{HCV} clustering algorithm, we briefly describe typical agglomerative clustering without considering the information from the geometry domain. For $i = 1,\ldots, n$, let $x_i(s_i) = (x_{i1}(s_i), \ldots, x_{ip}(s_i))'$ be a sample with p-dimensional attributes and $s_i$ is its location center on the geometry domain. The clustering  algorithm starts by setting each sample as its single-element cluster, and recursively merging the pair of clusters with the smallest dissimilarity.  Euclidean distance is an often used dissimilarity for pairs of samples on their $x$ values, i.e., 
\[
  d(x(s_i), x(s_j)) = \left(\sum^p_{k=1}\left|x_{ik}(s_i) - x_{jk}(s_j)\right|^2\right)^\frac{1}{2}.
\] 
It is easy to see at the $t$-th iteration, there will be $n-t+1$ candidate clusters among which a pair will be merged. We denote these  candidate clusters as $\{C_h^{t}, h=1,\ldots,n-t+1\}$. A cluster may contain more than one single point, and hence, several linkage methods for determining the between-cluster dissimilarity has been proposed. The well known methods, including single, complete, average, and Ward's methods are all follow the Lance-Williams formula with different linear coefficient \citep{lance1967general} in such a form:  
\begin{align}
\label{lw}
    d\left(C_h^{\{t\}}, C_k^{\{t+1\}}\right) &=  \alpha_i d\left(C_i^{\{t\}}, C_h^{\{t\}}\right) + \alpha_j d\left(C_j^{\{t\}}, C_h^{\{t\}}\right) + \beta d\left(C_i^{\{t\}}, C_j^{\{t\}}\right) \nonumber \\
    &+ \gamma \left|d\left(C_h^{\{t\}} , C_i^{\{t\}}\right) - d\left(C_h^{\{t\}}, C_j^{\{t\}}\right)\right|,
\end{align}
where in the $(t+1)$-th iteration $C_k^{\{t+1\}}$ is the cluster aggregated from cluster $C_i^{\{t\}}$ and cluster $C_j^{\{t\}}$. One applies a selected linkage method recursively to compute dissimilarity, and merge the closest pair of clusters, to find a hierarchy of clusters until all  samples are in a single cluster.

The idea of our HCV algorithm is generally based on the common edges of units on the geometry diagram. If any pairs of units have a common edge in the Voronoi diagram, then they are considered to have spatial proximity and geometry connectedness. In each iteration, the major difference between HCV and a typical  hierarchical agglomerative clustering algorithm is that, we only focus on the geometry-connected groups. The next two candidate groups to be aggregated must share a common edge. That is, a pair of clusters $C_h^{\{t\}}$ and $C_k^{\{t\}}$ can be merged in the $(t+1)$-th iteration only if $A^{\{t\}}_{hk}= 1$, where the stepwise updated adjacency matrix $A^{\{t\}}$ is recursively defined as
\begin{equation}
    A^{\{0\}}_{ij}=
\begin{cases}
0, & \text{if } i=j \text{ or the } i \text{-th and } j \text{-th smaples have no common edge;}\\
1, & \text{if the } i \text{-th and } j \text{-th samples have a common edge;}\\
\end{cases}
\label{ini-adj}
\end{equation}

\begin{equation}
\label{adj-update}
    A_{hk}^{\{t\}} = 
\begin{cases}
0, & \text{if } A_{hj}^{\{t-1\}} = 0 \text{ and } A_{hi}^{\{t-1\}} = 0,\\
1, & \text{if } A_{hj}^{\{t-1\}} = 1 \text{ or } A_{hi}^{\{t-1\}} = 1.\\
\end{cases}
\end{equation}
Putting these ideas altogether, the algorithm details of the implemented \texttt{HCV} function are given in Algorithm \ref{alg:HCV}.

There are several linkage algorithms have been provided in \texttt{HCV}, including "ward", "single", "complete", "average", "weight", "median", and "centroid". Default is 'ward'. Current available methods in \texttt{HCV} for the dissimilarity of the feature domain  are "euclidean", "correlation", "abscor", "maximum", "manhattan", "canberra", "binary" and "minkowski". Default is 'euclidean'.

\begin{algorithm}
\caption{Hierarchical Clustering from Vertex-links}
\begin{algorithmic}[1]
\State Input geometry\_domain, feature\_domain, and linkage.
\State Set $A^{\{0\}} \equiv$ the adjacency matrix  for geometry\_domain using (\ref{ini-adj}).  
\State Calculate $D^{\{0\}} \equiv$ the dissimilarity matrix for feature\_domain.
\State Set $\left\{C_i^{\{0\}}; i=1,\ldots,n \right\}$ with each sample being a single-element cluster.  
\State Set $t:=0$.
\State Create $\mathcal{T}$ as an edgeless graph for the initial tree structure.
\State Create Tree.Height as an empty array.  
\While{any off-diagonal element of $A^{\{t\}}$ is nonzero }
 \State Aggregate $C_i^{\{t\}}$ and $C_j^{\{t\}}$ 
 into $C_{NEXT}$ if $\displaystyle\mathop{\arg\min}_{i,j,A_{ij}=1}D_{ij}^{\{t\}}$
 \State Set $t:=t+1$. 
 \State Update $D^{\{t\}}$ based on (\ref{lw}) according to the input linkage.
 \State Update $A^{\{t\}}$ based on (\ref{adj-update}).
 \State Update Tree.Height[$t$] = $\displaystyle d\left(C_i^{\{t\}}, C_j^{\{t\}}\right)$.
 \State  Add $C_{NEXT}$ into $\mathcal{T}$ as a new node $C_k^{\{t+1\}}$.
 \State Draw edges from $C_k^{\{t+1\}}$ to its left and right nodes, i.e., $C_i^{\{t\}}$ and $C_j^{\{t\}}$.  
\EndWhile
\State \textbf{return} Tree.Height and $\mathcal{T}$ containing list of clusters and the hierarchy  
\end{algorithmic}
\label{alg:HCV} 
\end{algorithm}

\subsection{Key subroutine: Determining number of clusters}

Another crucial  issue in cluster analysis is to determine an appropriate number of clusters. There are several internal indices and external indices; see e.g. \citet{clusterCrit}.   However, these indices consider only the within-cluster sum of squared difference of attributes on the feature domain, which overlooks the information on the geometry domain. In view of this, we propose a novel internal index named  \emph{Spatial Mixture Index} (SMI) for determining the  number of clusters. 

In some applications, having the number of clusters  is not satisfactory enough. We may want to know the stability of clusters and their adherent members. To this end, we provide another method `M3C' based on Monte Carlo simulations to find stable clusters. 

Both `SMI' and `M3C' are available methods in the wrapper function \texttt{getCluser}, which provides users a more convenient tool for  the  number of clusters.

\subsubsection{SMI} \label{SMI}

By connecting the pairs of points sharing common edges on the geometry domain, we have many paths between samples. Denote $\ell_{ij}$ the minimum number of paths through which $s_i$ can reach $s_j$.   
SMI is an internal index involving the property of spatial proximity by considering the edge length of polygons or of triangles after a tessellation. 
SMI is defined in (\ref{smie}), where ${\Delta}_{k}$ and $e_k$ represent the average squared distance within the $k$-th cluster for  the feature domain and the geometry domain, respectively. The overall heterogeneity is a weighted sum of both squared distances for all clusters, with weights depending on $f(\cdot)$.

\begin{equation}
    \label{smie}
    \mbox{SMI}(K)= \frac{1}{K}\sum_{k=1}^K \left[f(\alpha_k){\Delta}_{k} +  (1 - f(\alpha_k)) e_k \right],
\end{equation}
where     
\begin{align*}
    {\Delta}_{k} &=\frac{1}{|C_k|^2-|C_k|}\sum_{i \in C_k}\sum_{j \in C_k \backslash{i} }  d(x_i(s_i),x_j(s_j))^2,\\
    e_k &= \frac{\rho}{|C_k|^2-|C_k|}\sum_{i \in C_k}\sum_{j \in C_k \backslash{i} } \ell_{ij}^2, \\       
    \rho &= \frac{\sum_{i=1}^n \sum_{j=1}^{i-1} d(x_i(s_i),x_j(s_j))^2}{\sum_{i=1}^n \sum_{j=1}^{i-1}\ell_{ij}^2},    
\end{align*}
and  $\alpha_k$ and the function $f$ are defined as follows: 

\begin{equation*}
    \begin{aligned}
    \alpha_k &= \frac{Ke_k - \sum_{j=1}^Ke_j}{\sum_{j=1}^Ke_j},\\
    f(x) &= (1 + e^{-x})^{-1}.
\end{aligned}
\end{equation*}
The squared distances are analogous to within-cluster sum of squares, and $\rho$ deals with potiential different scales between ${\Delta}_{k}$ and $e_k$. 
 We suggest to use $K^*=\mbox{arg min}_K \mbox{SMI}(K)/\mbox{SMI}(K-1)$ for selecting the number of clusters.   
Our \texttt{getCluster} fucntion has a \texttt{method} argument. If  \texttt{method = `SMI'} is specified, values of SMI
from 2 to \texttt{Kmax} clusters will be compared, and 
individual cluster members are reported, under the selected  number of clusters based on SMI.

\subsubsection{M3C}
We modified a method in the \pkg{M3C} package \citep{M3Cbib} to take the uncertainty of cluster into account via Monte Carlo simulations and calculating the consensus
matrix.  \pkg{M3C} compares the result of the clusters at hand with the clusters generated by simulations, and it determines
the appropriate number of clusters based on the most significant proportion of ambiguous clustering (PAC) score.
As its design is not for spatial data clustering, \pkg{M3C} does not consider the information on the geometry domain.
We construct a geometry embedded dissimilarity matrix (GEDM) to fill the gap, and the procedure is explained in the following.

Our method utilizes the output of the \texttt{HCV} algorithm, which carries information of the dissimilarity between clusters and the constraints on the geometry.  We turn such information into a GEDM through Algorithm \ref{GEDM-alg}. Then the GEDM is converted into Euclidean coordinates using a metric multidimensional scaling, followed by applying Monte Carlo reference-based consensus clustering of the \pkg{M3C} package. The default score to evaluate cluster stability is PAC score, but users can specify an alternative score  in the \texttt{criterion} argument when necessary.  

Similar to that with \texttt{method = `SMI'}, if  \texttt{method = `M3C'} is used in the implemented \texttt{getCluster} fucntion,
individual cluster members are reported, under the selected  number of clusters having the smallest score.

\begin{algorithm}\caption{Geometry Embedded Dissimilarity Matrix}
\begin{algorithmic}[1]
    \State Input an \texttt{HCV} object with Tree.Height and $\mathcal{T}$.
    \State Set I$_{max}$ the length of Tree.Height in the input \texttt{HCV} object
    \State Create a matrix $M$ with zero diagonals and all off-diagonals being Tree.Height[I$_{max}$].
    \State Set GEDM $:= M$. 
    \For{$t$ from I$_{max}$ to 1} 
    \State LeftMembers := members in LeftNode of $C_k^{\{t\}}$ in $\mathcal{T}$
    \State RightMembers := members in RightNode of $C_k^{\{t\}}$ in $\mathcal{T}$       
        \For{$\ell$ in LeftMembers}
            \For{$r$ in RightMembers}
                \State GEDM[$\ell$, $r$] = $\min$(Tree.Height[$t$], GEDM[$\ell$,$r$])
                \State GEDM[$r$, $\ell$] = GEDM[$\ell$, $r$]
            \EndFor
        \EndFor
    \EndFor
    \State \textbf{return} GEDM
\end{algorithmic}
\label{GEDM-alg}
\end{algorithm}

\section{Simulation Examples and Real Data}

\subsection{Simulation data}

We used the way to generate synthetic point-level data propose by \citet{lin2005dual} as follows. The procedure is also implemented as the \texttt{synthetic\_data} function in our package.  

\begin{enumerate}
    \item Specify the number of clusters $(k)$, number of points $(n)$, the dimension of feature domain $(p_1)$ and geometry domain $(p_2)$.
    \item Select exactly $k$ points from the domain $(0,1)^{p_1}$ randomly as the center of the $k$ clusters on feature domain, i.e., $\mu^{feat}_1,\ldots, \mu^{feat}_k$. 
    \item Select exactly $k$ points from the domain $(0,1)^{p_2}$ randomly as the center of the $k$ clusters on geometry domain, i.e., $\mu^{geo}_1,\ldots, \mu^{geo}_k$ 
    \item Draw a random point $o$ from joint uniform distribution over the geometry domain.    
    \item Assign point $o$ to $C_i$ by the probability mass function (PMF) as  $P(o \in C_i) = \frac{d\left(o,\mu^{geo}_i\right)^{-f}}{\sum_{j=1}^k {d\left(o,\mu^{geo}_j\right)^{-f}}}$
    \item Draw a sample form joint normal distribution $N(\mu_i^{feat}, r^2I)$ as the feature domain attribute for the point $o$ in Step 4.
    \item Repeat Step 4 to 6 until $n$ points have been sampled.
\end{enumerate}

In the above process, $C_i^{feat}, C_i^{geo}$ are the centers of feature domain and geometry domain of the cluster $C_i$ respectively. It is easy to see that the parameter $f$ controls the degree of spatial homogeneity in step 5. 
the larger $f$ value presents, the more spatial homogeneity has been introduced, and we controls the variation of the feature domain attribute by parameter $r$. 

Here we present an example for clustering of point-level data. The parameters of synthetic data were $(k, f, r, n, p_1, p_2) = (3,30,0.02,300,2,2)$.

\begin{example}
set.seed(0)
pcase <- synthetic_data(3,30,0.02,300,2,2)
labcolor <- (pcase$labels+1)
par(mfrow = c(1,2))
plot(pcase$feat, col = labcolor, pch=19, xlab = 'First attribute', 
     ylab = 'Second attribute', main = 'Feature domain')
plot(pcase$geo, col = labcolor, pch=19, xlab = 'First attribute', 
     ylab = 'Second attribute', main = 'Geometry domain')
\end{example}

\begin{figure}[H]
\centering
\includegraphics[width=1\textwidth]{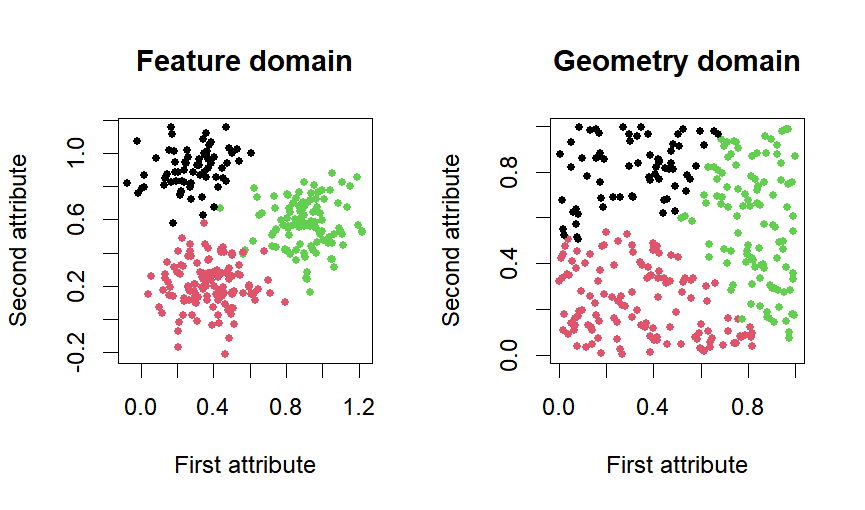}
\caption{The feature domain and the geometry domain of the simulated point-level data, where distinct colors stand for different clusters.}
\label{syn}
\end{figure}

\begin{figure}
\centering
\includegraphics[width=0.6\textwidth]{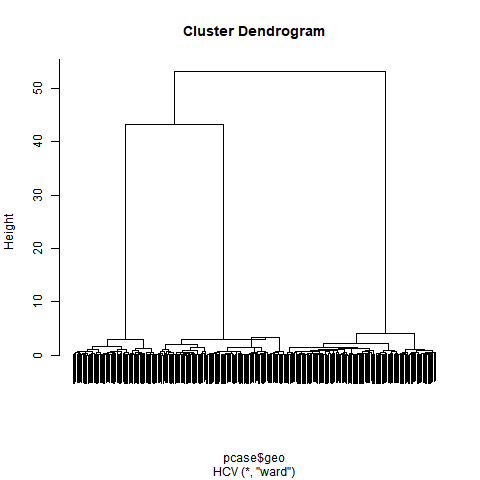}
\caption{ The cluster dendrogram for the simulated data on Figure \ref{syn} using \texttt{HCV}.}
\label{tree}
\end{figure}

\begin{figure}
\centering
\includegraphics[width=1\textwidth]{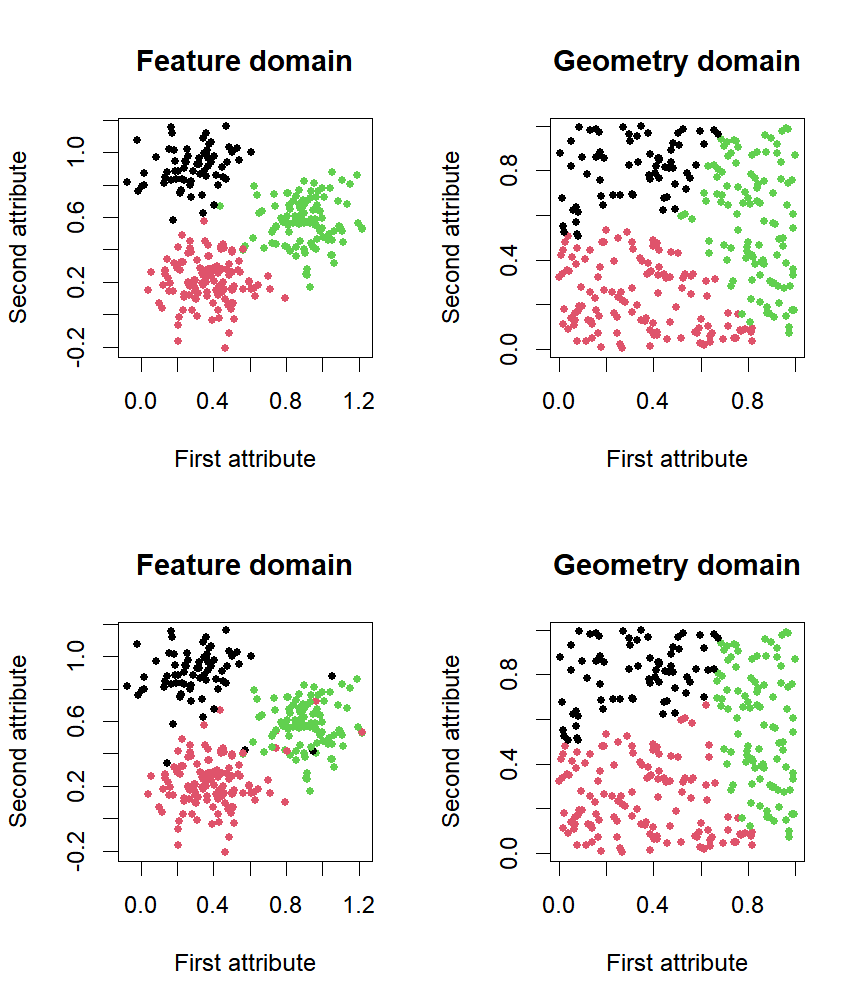}
\caption{The scatterplot with found clusters using the HCV algorithm (top panel) and using the \pkg{SPODT} package (bottom panel), where distinct colors stand for different clusters in each method.}
\label{syn-cluster}
\end{figure}

The simulated data are shown in Figure \ref{syn}. We then compare the results from our package using \texttt{method}='SMI' with that from the \pkg{SPODT} package. The detailed codes are given below. Since \texttt{HCV} outputs an \texttt{hclust} object, we can easily plot the dengrogram as Figure \ref{tree}shown, and  use \texttt{cutree} to form final clusters. However, a closer examination will reveal that crossovers or inversions occur on such a dendrogram, and hence \texttt{cutree} should be applied with care. We recommend  \texttt{getCluster} instead for determining clusters. The found clusters are shown in  Figure \ref{syn-cluster}.
As can be seen, the proposed method almost recovers the true clusters, while there are more mismatches based on \pkg{SPODT}. 

\begin{example}
# clustering with HCV
HCVobj <- HCV(pcase$geo, pcase$feat)
smi <- getCluster(HCVobj,method="SMI")
plot(HCVobj)

# clustering with SPODT
library(cluster)
library(SPODT)
grp <- pam(pcase$feat,k=3)
dataset <- data.frame(pcase$geo,grp$clustering) 
colnames(dataset) <- c('x1','x2','grp')
coordinates(dataset) <- c("x1","x2")
spc <- spodt(dataset@data$grp~1, dataset, graft = 0.3)  

# plots for results
par(mfrow=c(2,2))
plot(pcase$feat, col=factor(smi),pch=19, xlab = 'First attribute',
     ylab = 'Second attribute',main = 'Feature domain')         
plot(pcase$geo, col=factor(smi),pch=19, xlab = 'First attribute',
     ylab = 'Second attribute',main = 'Geometry domain') 
plot(pcase$feat, col=factor(spc@partition),pch=19, xlab = 'First attribute',
     ylab = 'Second attribute',main = 'Feature domain')         
plot(pcase$geo, col=factor(spc@partition),pch=19, xlab = 'First attribute',
     ylab = 'Second attribute',main = 'Geometry domain')            
\end{example}

\subsection{Real application: French municipalities data}

\begin{figure}
\centering
\subfigure[City employee rate]{
\includegraphics[width=0.45\textwidth]{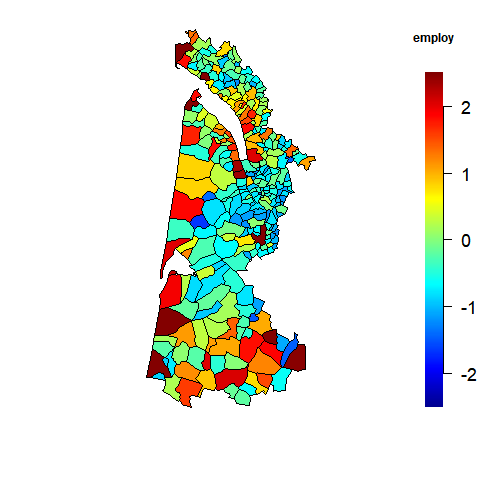}
}
\quad
\subfigure[City graduation rate]{
\includegraphics[width=0.45\textwidth]{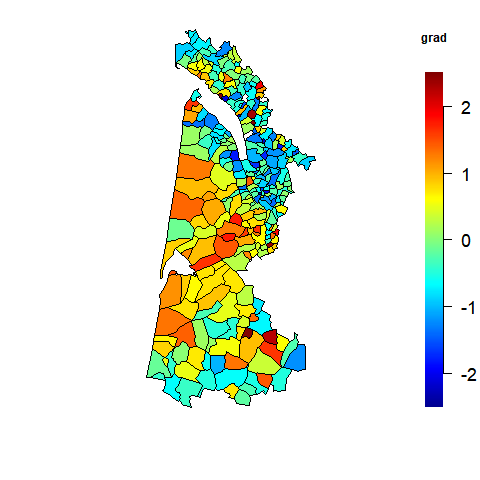}
}
\quad
\subfigure[Ratio of apartment housing]{
\includegraphics[width=0.45\textwidth]{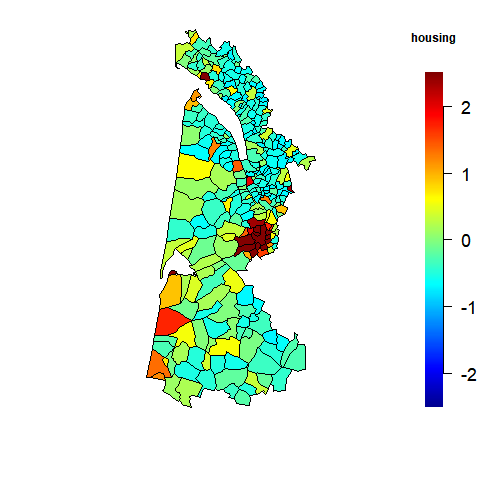}
}
\quad
\subfigure[Agricultural area]{
\includegraphics[width=0.45\textwidth]{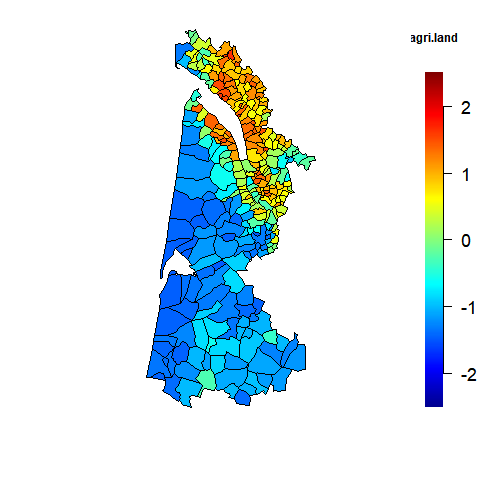}
}
\caption{Heatmaps for standardized values of the four socio-economic variables in the \texttt{estuary} data.}
\label{heatmap}
\end{figure}

We used the \texttt{estuary} data in the \pkg{ClustGeo} package, which involved 303 France cities located along Gironde estuary. There are 
four attributes on the feaure domain: city employee rate, city graduation rate, ratio of apartment housing and agricultural area, and these are areal data on the geometry domain. For understanding the spatial distribution of the four features, we visualize via heatmaps and represent the magnitude of standardized features in Figure \ref{heatmap}. It is obvious that there are several potential outliers in city employee rate and  ratio of apartment housing that have extremely large values.

\begin{example}
library(ClustGeo)
library(fields)
data(estuary)
colors <- tim.colors(303)
lim <- c(-2.5,2.5)
dat <- estuary$dat
plotMap(estuary$map, scale(dat[,1]), color=colors, zlim=lim, bar='employ')
plotMap(estuary$map, scale(dat[,2]), color=colors, zlim=lim, bar='grad')
plotMap(estuary$map, scale(dat[,3]), color=colors, zlim=lim, bar='housing')
plotMap(estuary$map, scale(dat[,4]), color=colors, zlim=lim, bar='agriland')
\end{example}

\begin{figure}
\centering
\includegraphics[width=1\textwidth]{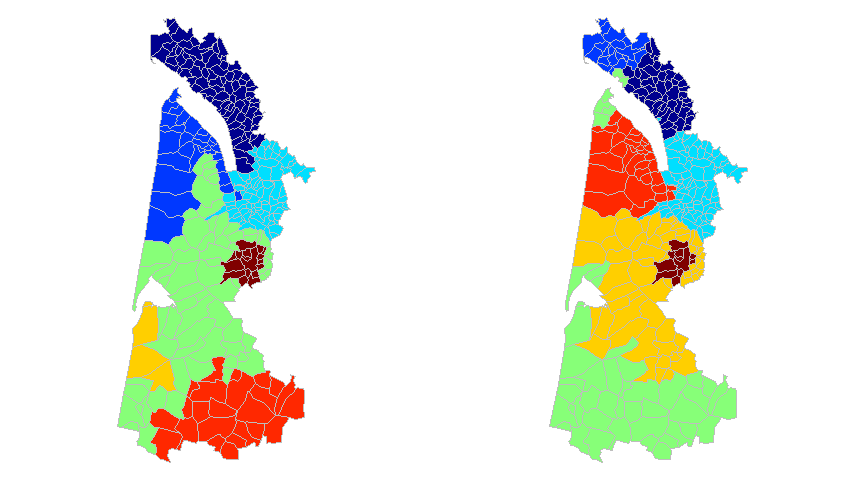}
\caption{The found clusters using the HCV algorithm (left) and using the \pkg{ClustGeo} package (right),  where distinct colors stand for different clusters in each method.}
\label{estuary-cluster}
\end{figure}

In this example, we present to input a user-specified dissimilarity matrix and adjacency matrix, by setting \texttt{diss = 'precomputed'} and \texttt{adjacency = TRUE} as the codes given below. We use squared Euclidean distance as the dissimilarity, and covert the \texttt{SpatialPolygonsDataFrame} object \texttt{estuary\$map} to an adjacency matrix. After calling the \texttt{HCV} function, we apply  the \texttt{getCluster} function with \texttt{method='M3C'} to find the most stable clusters. We also compare it with the clusters found by \pkg{ClustGeo}.  The result are shown in Figure \ref{estuary-cluster}. Some cluster patterns are quite similar in the two methods, although \pkg{ClustGeo} has more spatially fragmented clusters. If spatial contiguity is of interest, the HCV algorithm is preferred.

\begin{example}
# clustering with HCV
feat <- as.matrix(scale(dat))
fdist <- as.matrix(dist(feat)) ** 2
geo <- estuary$map
adj <- rgeos::gTouches(geo, byid=TRUE) * 1
tree1 <- HCV(adj, fdist,  adjacency = T, diss = 'precomputed')
m3c <- getCluster(tree1, method='M3C')

# clustering with ClustGeo
D0 <- dist(scale(dat))
D1 <- as.dist(1-adj)
tree2 <- hclustgeo(D0,D1,alpha=0.4)
cg <- cutree(tree2,7)

# plots for results
par(mfrow=c(1,2),mar=c(0,0,0,0))
plot(geo, col = tim.colors(7)[m3c], border='grey')
plot(map,border="grey",col=tim.colors(7)[cg])
\end{example}

%
%
%

\section{Summary}
In this paper, the R package \pkg{HCV} is presented for spatial data clustering 
with the contiguity property. We modified hierarchical agglomerative clustering
algorithms and methods for determining the number of clusters, such that the spatial information on the geometry domain can be reasonablly incorproated.  
We demonstrate the package applications to point-level data and areal data.  By these examples, we show that \pkg{HCV} is a unified framework for clustering both types of spatial data.


\section{Acknowledgements}
This research was supported in part by the Ministry of Science and Technology, Taiwan under Grants 110-2628-M-110-001-MY3. We also thank Yung-Jie Ye for 
part of the initial code prototype.

\bibliography{tzeng-hsu}

\address{ShengLi Tzeng\\
Department of Applied Mathematics\\
National Sun Yat-sen University\\
Kaohsiung, Taiwan, 80424 \\
  \email{slt.cmu@gmail.com}}

\address{Hao-Yun Hsu\\
Department of Applied Mathematics\\
National Sun Yat-sen University\\
Kaohsiung, Taiwan, 80424 \\
  \email{zoez5230100@gmail.com}}